\documentclass{arxiv}

\usepackage{algorithm}
\usepackage{algorithmic}
\usepackage{amssymb}
\usepackage{graphicx}
\usepackage{subfigure}
\usepackage{epstopdf}
\usepackage{color}
\usepackage{url}
\font\msbm=msbm10 at 10pt
\newcommand{\ZZ}{\mbox{\msbm Z}}

\def \Z {{\ZZ}}

\begin{document}
%\mainmatter  % start of an individual contribution
\title{Modular Arithmetic Expressions and Primality Testing via DNA Self-Assembly}
%\title{XTile: Modular Arithmetic Package for DNA Self-Assembly}
%\titlerunning{Modular Arithmetic Expressions and Primality Testing}
%\author{}
\author{Abhishek Chhajer, Manish K. Gupta,   Sandeep  Vasani,  Jaley Dholakiya}
%
%\authorrunning{}
\affiliation{Laboratory of Natural Information Processing,\\
Dhirubhai Ambani Institute of Information and Communication Technology,\\
Post Bag Number 4 , Near Indroda Circle,\\
Gandhinagar, Gujarat   382007, India\\
\email{ m.k.gupta@ieee.org }\\
%\email{ @}\\
\url{http://www.guptalab.org}}
%\maketitle              % typeset the title of the contribution
%\pagestyle{empty}
%\thispagestyle{empty}
%\vspace*{0.5cm}
%\vspace*{1cm}
\begin{abstract}
{\small Self-assembly is a fundamental process by which supramolecular species form spontaneously from their components. This process is ubiquitous throughout the life chemistry and is central to biological information processing.  Algorithms for solving many mathematical and computational problems via tile self assembly has been proposed by many researchers in the last decade. In particular tile set for doing basic arithmetic of two inputs have been given. In this work we give tile set for doing basic arithmetic (addition, subtraction, multiplication) of $n$ inputs and subsequently computing its modulo.  We also present a tile set for primality testing.  Finally  we  present a software 'xtilemod' for doing modular arithmetic. This simplifies the task of creating the input files to xgrow simulator for doing basic (addition, subtraction, multiplication and division) as well as modular arithmetic of $n$ inputs.} Similar software for creating tile set for primality testing is also given. 
\end{abstract}
%\vspace{1.5cm} 
\vspace*{0.5cm}
%%%%%%%%%%%%%%%%%%%%%%%%%%%%%%%%%%%%%%%%%
{\it Keywords: }{\small DNA self-assembly, error-correction, algorithmic self-assembly,  Wang tile, DNA computing, Xgrow, XTile} 
%\vspace{0.5cm}
%\newpage
%%%%%%%%%%%%%%%%%
\section{\large Introduction}
\normalsize Self-assembly is a natural phenomenon observed at many places in nature such as formation of galaxies, formation of coral reefs,  crystal growth etc. In $1996$ Erik Winfree of California Institute of Technology, showed that it can be used to perform nano-scale computations ~\cite{4281288}. This paved way for the birth of algorithmic self-assembly utilizing knowledge of three fields - DNA Nanotechnology~\cite{rls01} (due to the pioneering work of Ned Seeman in $1980s$), DNA Computing (due to the pioneering work of L. Adleman in $1994$) and Tiling Theory ~\cite{wang63} (due to pioneering work of H. Wang who showed that zigsaw shaped colored tiles can simulate universal Turing machine). Winfree formulated the idea of molecular Wang tile using all this and showed that it can simulate universal Turing machine using abstract Tile Assembly Model (aTAM) ~\cite{WIN1,WIN2}.  Algorithms for solving many mathematical and computational problems via tile self assembly has been proposed by many researchers in the last decade ~\cite{Brun08np-c,Brun08PhD,Brun08factor,4656700}. The addition of two numbers with Wang tiles is given in the book ~\cite{gs87}. In $2006$ Brun ~\cite{Brun07fnano,Brun07arith} gave algorithm for addition and multiplication of two numbers with $aTAM$ and  in $2008$ Zhang and his coworkers  ~\cite{Zhang2009377} gave algorithm for subtraction and division of two numbers in $aTAM$. In this work, we extend the results and provide tile set for  addition and multiplication of $n$ numbers together with a tile set for arithmetic expression involving addition and subtraction. This is further extended to a tile set for modular arithmetic expression.  Winfree's tile assembly model can be simulated using a program developed by him called Xgrow ~\cite{WIN3}.  The input to Xgrow is  a .tiles file. Many such standard  .tiles files are distributed with the Xgrow package. In this paper, we also present a software package 'Xtilemod' for modular airthmetic  that can be used to create such input files for the Xgrow simulator of Winfree by providing the basic arithmetic expressions. For further details on algorithmic self-assembly the reader is referred to excellent papers ~\cite{wibe04,rsy04,chgo04} and thesis~\cite{win98,Brun08PhD}.

This paper is organized as follows. In Section $2$, we give a brief overview of the aTAM.  Section $3$ provides the algorithms for addition and multiplication of $n$ numbers via $aTAM$. Algorithm for arithmetic expression of $n$ numbers involving  addition and subtraction is described in Section $4$ and algorithm for modular arithmetic  expression of $n$ numbers involving addition and subtraction is given in Section $5$. Section $6$   gives tile set for primality testing. In Section $7$ we describe a software for modular arithmetic  expressions which can produce the input to xgow simulator. Finally Section $8$  concludes the paper with some general remarks.
%%%%%%%%%%
\section{Background}
\begin{figure}
\centering
\includegraphics[scale=.35]{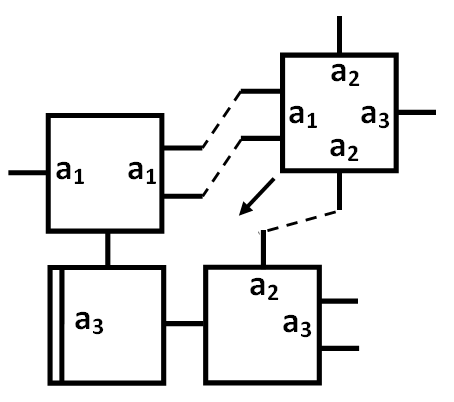}\\
%\centerline{\epsfxsize\hsize\epsfbox{RNA.eps}}
  \caption{abstract Tile Assembly Model (aTAM) : A Tile attaches to a Seed Configuration of 3 Tiles}\label{aTAM}
\end{figure}

Self assembly of DNA molecules can be approximated by mathematical models. Several such models have been given by many authors. In ~\cite{WIN1}, Winfree gave a model called aTAM. The basic building 
block of a tile assembly model are square tiles. Each tile has $4$ edges viz north, east, south and west with glues assigned to each edge. Each glue has some strength (usually an integer). Assembly starts with some seed tiles and frame tiles giving the initial input of the aTAM and a tile can attach to another tile if (1) the edges have matching glues and (2) the combined glue strength of  both the tiles is greater than or equal to system temperature (again an integer, usually denoted by $\tau$ and we take $\tau=2$). Tiles can not be rotated. Under these settings system will grow from some initial tile set configuration to form a pattern or depending upon problem input it solves a problem. In Figure \ref{aTAM},  a tile attaches to a seed configuration of 3 tiles. East glue $a_1$ of configuration tile and west glue of outside  tile is of strength $2$, all other tiles have  glues of strength $1$ except north of configuration tile and south of each tile in seed configuration that has strength $0$. System temperature is $\tau=2$. It is shown by Winfree that such system is Turing universal. For more formal description the reader is referred to ~\cite{Brun08DNA-lncs,Brun07fnano,Brun07arith}. 
%See Figure~\ref{fig:bcounter} for a tile set for binary counting. 
%\begin{figure}[ht!]
    %\label{fig:subfigures}
    %\begin{center}
%
        %\subfigure[Counter]{%
            %\label{fig:bincount}          
%$
%\begin{array}{cccccccccc}
   %&\vdots  &\vdots  &\vdots  &\vdots&&&& \\ 
%\ldots  & 0 & 1 & 1 & 0&&&&\\
%\ldots  & 0 & 1 & 0 & 1&&&&\\
%\ldots  & 0 & 1 & 0 & 0&&&&\\
%\ldots  & 0 & 0 & 1 & 1&&&&\\
%\ldots  & 0 & 0 & 1 & 0&&&&\\
%\ldots  & 0 & 0 & 0 & 1&&&&\\
%\ldots  & 0 & 0 & 0 &0&&&&
%\end{array} 
%$
%            \includegraphics[width=0.4\textwidth]{theorem1/nsum8compu_a}
    %    }%
  %\\      
      %  \subfigure[Binary Counter Tiles and their Growth]{%
          % \label{fig:bcounter}
           %\includegraphics[width=0.90\textwidth]{Binarycounter.eps}
        %}\\ %  ------- End of the first row ----------------------%
       %
    %\end{center}
    %\caption{%
      %Figure  \ref{fig:bincount} is an unbounded binary counter pattern and \ref{fig:bcounter} shows how 4 computational tiles yields the binary counter pattern with gray color tile representing "1" and white color tile representing  "0".
     %}%
%\end{figure}
%\begin{figure}
%\centering
%\begin{center}
%$
%\begin{array}{ccccc}
   %&\vdots  &\vdots  &\vdots  &\vdots \\ 
%\ldots  & 0 & 1 & 1 & 0\\
%\ldots  & 0 & 1 & 0 & 1\\
%\ldots  & 0 & 1 & 0 & 0\\
%\ldots  & 0 & 0 & 1 & 1\\
%\ldots  & 0 & 0 & 1 & 0\\
%\ldots  & 0 & 0 & 0 & 1\\
%\ldots  & 0 & 0 & 0 & 0
%\end{array} 
%$
%\end{center}
%\includegraphics[scale=.32]{Binarycounter.eps}\\
%\centerline{\epsfxsize\hsize\epsfbox{RNA.eps}}
  %\caption{Binary Counter}\label{bcounter}
%\end{figure}
Recently basic arithmetic of two inputs with DNA Self Assembly have been studied by many authors \cite{Brun07fnano,Brun07arith,Zhang2009377}. In particular, two algorithms have been proposed by Brun \cite{Brun07fnano,Brun07arith} for adding two numbers. Algorithm using $L$-configuration uses $16$ computational tiles for addition whereas there is another economical tile set for adding two numbers which takes just $8$ computational tiles. In ~\cite{Brun07arith}, Brun also gave algorithm for multiplication of two numbers with $28$ computational tiles. Algorithm for subtraction and division are given in \cite{Zhang2009377}.  In the next sections we will assume that the reader is familiar with the notations and definitions of ~\cite{Brun07arith,Zhang2009377}. 
%\section{Basic Arithmetic of two inputs with DNA Self Assembly}
%\subsection{Addition of two Numbers using $L$ Configuration with $16$ Tiles}
%\subsection{Addition of two Number with $8$ Tiles}
%\subsection{Addition of two Numbers using $L$ Configuration with variable number of Tiles}
%\subsection{Multiplication of two Numbers with $28$ Tiles}
%\subsection{Subtraction of two Numbers with  Tiles}
%\subsection{Division of two Numbers  with  Tiles}
\section{Basic Arithmetic of $n$ inputs with DNA Self Assembly}
\subsection{Addition of $n$ Numbers with $8$ Computational Tiles}
\begin{figure}[htdp]
   % \label{fig:nsum8tiles}
    \begin{center}
        \subfigure[Typical Computational Tile, where $ cout= (a \wedge b) \vee (b \wedge c) \vee (c \wedge a) $]{%
            \label{fig:nsum8tilesa}
            \includegraphics[width=0.4\textwidth]{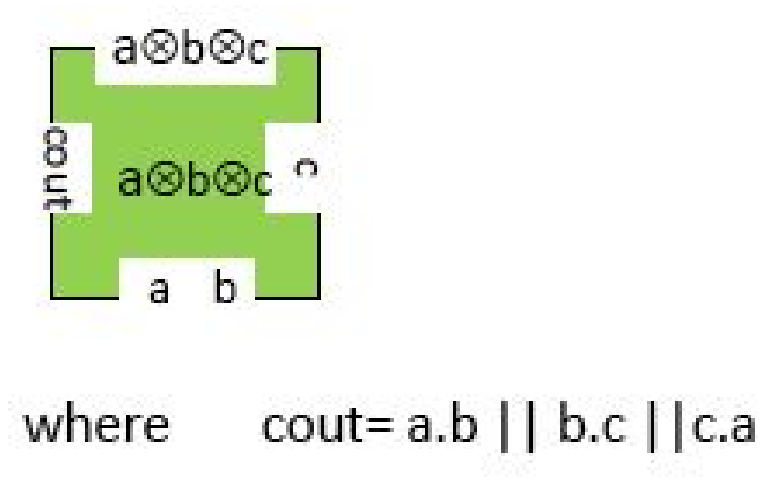}
        }%
        \subfigure[All 8 Computational Tiles]{%
           \label{fig:nsum8tilesb}
           \includegraphics[width=0.60\textwidth]{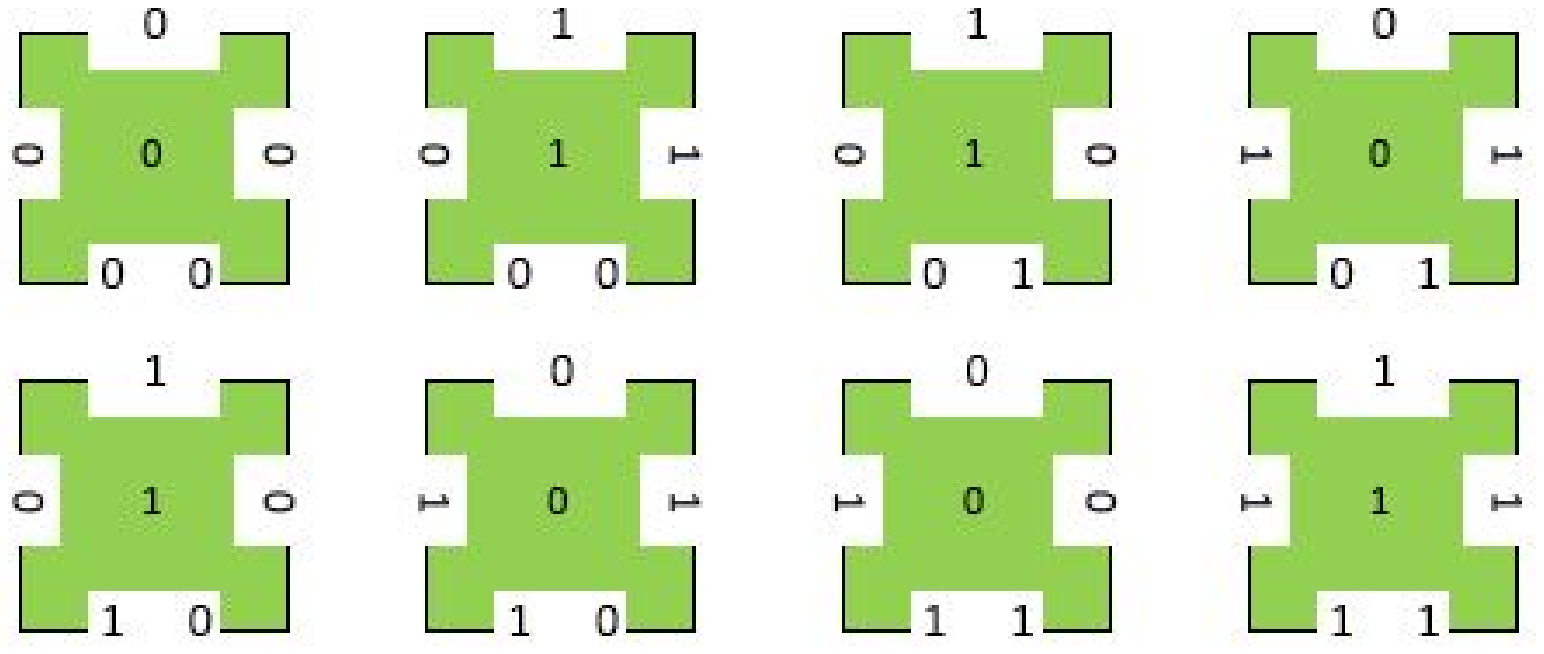}
        }\\ %  ------- End of the first row ----------------------%
        \subfigure[Two Top Frame Tiles]{%
            \label{fig:nsum8tilesc}
            \includegraphics[width=0.4\textwidth]{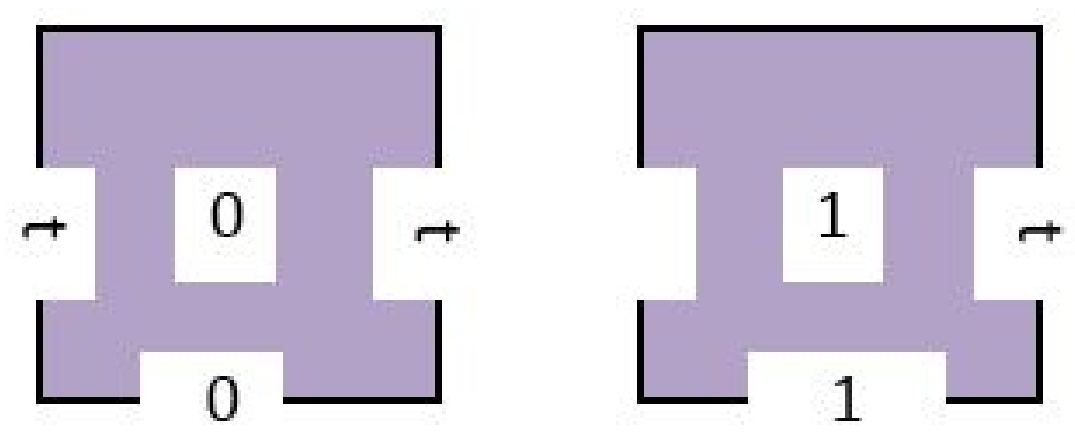}
        }%
          \hspace{.25in}
        \subfigure[All 4 Corner Tiles]{%
            \label{fig:nsum8tilesd}
            \includegraphics[width=0.65\textwidth]{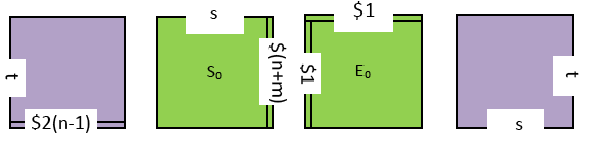}
        }\\%
        \subfigure[Left Frame Tiles \newline 
        ($1 \leq i \leq n-2$)]{%
            \label{fig:nsum8tilese}
            \includegraphics[width=0.4\textwidth]{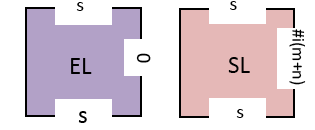}
        }%
        \hspace{.25in}
        \subfigure[Right Frame Tiles ($ER: 1 \leq i \leq n-1)$ \mbox{and}  ($SR: 1 \leq i \leq n-2)$ $\;\mbox{and}\;  \lambda=1+(i-1)(n+m)$]{%
           \label{fig:nsum8tilesf}
           \includegraphics[width=0.35\textwidth]{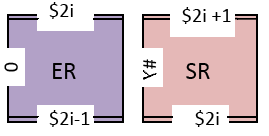}
        }\\ %  ------- End of the first row ----------------------%
        \subfigure[Lower Input Tile Types Encoding Two Input Bits) \newline ($1 \leq i \leq n+m-1)$  ]{%
            \label{fig:nsum8tilesg}
            \includegraphics[width=.38\textwidth]{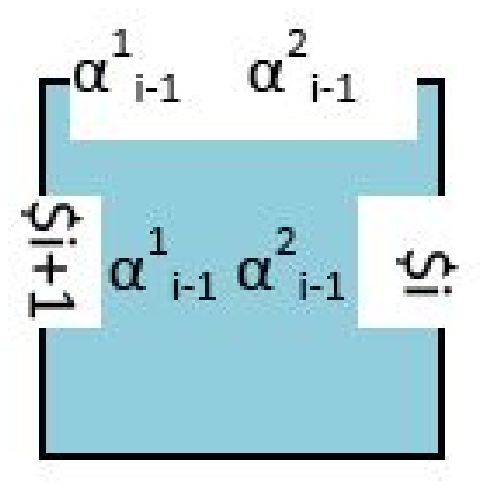}
        }%
         \hspace{.3in}
        \subfigure[Intermediate Input Tiles Types ($1 \leq i \leq n+m-1), 3 \leq j \leq n$ $\;\mbox{and}\; \lambda=i+(j-3)(n+m)$ ]{%
            \label{fig:nsum8tilesh}
            \includegraphics[width=0.5\textwidth]{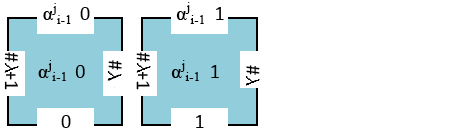}
        }% 
       %  \subfigure[Caption of Fourth Figure]{%
           % \label{fig:fourth}
            %\includegraphics[width=0.5\textwidth]{theorem1/nsum8tilesexample}
        %}%       
%
    \end{center}
    \caption{%
   Tile Set for Computing  $\sum_{i=1}^{n} a_i$ using $8$ Computational Tiles. 
   Note  that in Input Tiles (Figure \ref{fig:nsum8tilesg} and \ref{fig:nsum8tilesh}) each  $\alpha^{j}_{\theta} \in \Z_2$ also $a,b,c \in \{0,1\}$.  \label{fig:nsum8tiles}
     }%
\end{figure}

\begin{figure}
\centering
  %Requires \usepackage{graphicx}
\includegraphics[scale=.29]{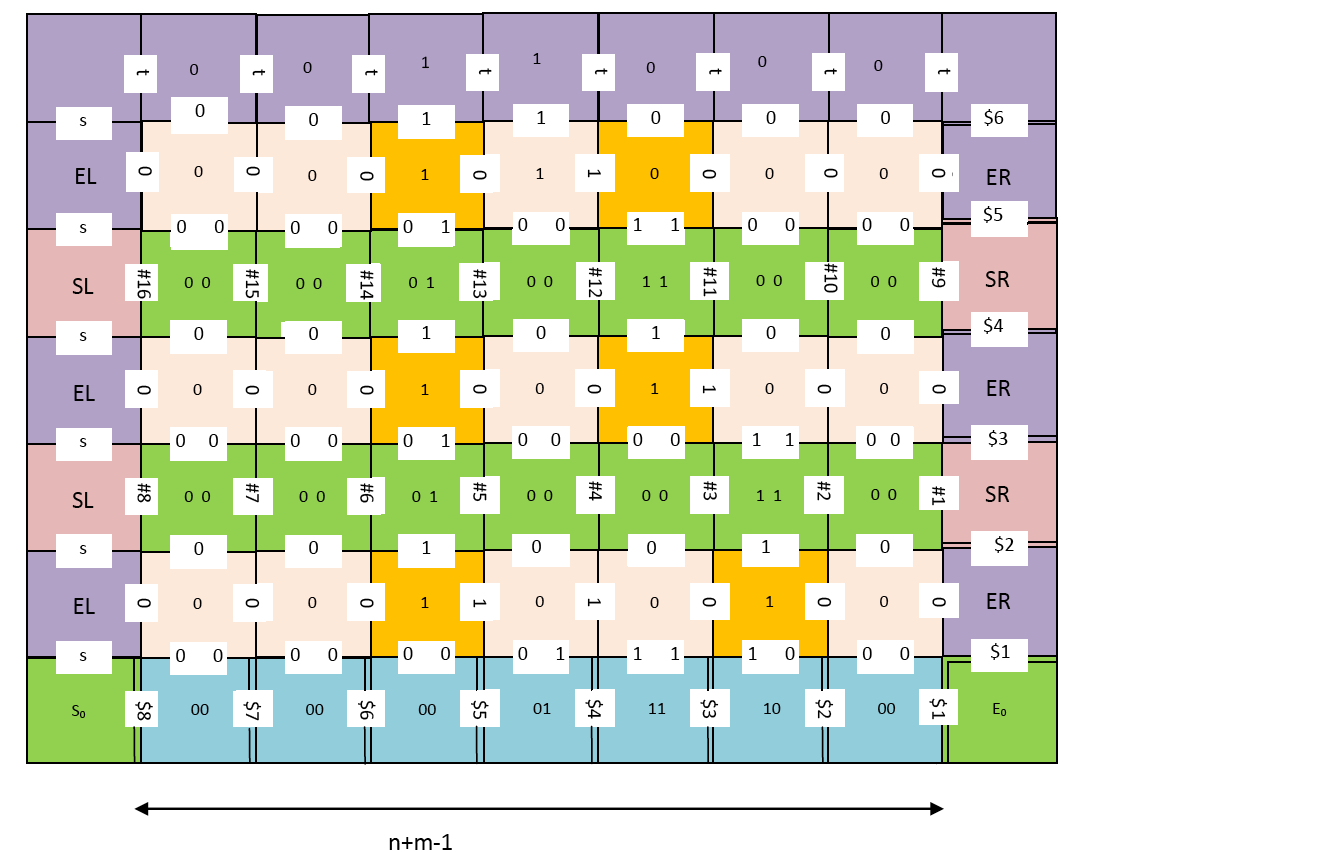}
%\includegraphics[scale=.5]{layeredseedm.eps}\\
%\centerline{\epsfxsize\hsize\epsfbox{RNA.eps}}
  \caption{Tiles Growth for Expression $12+6+2+4=24$ using $8$ Tile Types Configuration }\label{num8example}
\end{figure}

Suppose we want to add $n$ positive integers $\{ a_1, a_2, \ldots, a_n | a_i \in \Z^{+} \}$. We can write their binary representation as $a_j = \sum_{\theta =0}^{m_j} \alpha^{j}_{\theta} 2^{\theta}$, where $m_j =|a_j|$ is the size of the integer $a_j$ and $\alpha^{j}_{\theta} \in \Z_2$.  If $m = \max \{ m_j | 1 \leq j \leq n \}$ then we can pad extra zeros to write $a_j = \sum_{\theta =0}^{m} \alpha^j_{\theta} 2^{\theta}$. The next two Theorems  gives the tile set for computing $\sum_{i=1}^{n} a_i.$
\begin{theorem}
Let 
\[
\begin{array}{c}
\Sigma_{1}  =  \{s,t, 0, 1, 00, 01, 10, 11\} \cup \{ \#i | 1 \leq i  \leq  (m+n)(n-2)  \} 
\\
\cup \{ \$i | 1 \leq i \leq \max{(m+n,2n-2)} \}
\end{array}
\]
be the set of glues and let $T_1$  be a set of tiles  (different tiles types of $T_1$ are given in Table \ref{nsum8tiles} ) over  $\Sigma_1$  as described in Figure \ref{fig:nsum8tiles} then $T_1$  computes the function $\sum_{i=1}^{n} a_i$.
\end{theorem}
%\begin{proof} (Sketch) 
The logic of the system is identical to a series of one-bit full adders. Each solution tile takes a bit from each of the inputs on the south side and a carry bit of the previous solution from  the east side, and outputs the next carry bit on the west side and the sum on the north side.  After computing the sum of two numbers, we need to combine this sum with next input bitwise.  This process continues till all the integer inputs are added. To handle overflow we make size of each input $a_i$ as $n+m-1$ bits. We will illustrate it by an example. Figure \ref{num8example} shows a sample addition of $12+6+2+4=24$ ($m=4, n=4$). Input $12= {1100}_2$ (size $4$) and $6={110}_2$ (size $3$) are given at the bottommost (first)  row by adding extra zeros  (to adjust overflow we make size of each $n+m-1=7$ bits) and reading $12$ from right side and $6$ from left side. The sum is computed ($12+6=18={10010}_2$) in second row and then passed to the $3^{rd}$ row from left side, next input $2={10}_2$ is given from right side of $3^{rd}$ row and the sum is computed in $4^{th}$ row ($12+6+2=20={10100}_2$) which is then passed to the $5^{th}$ row from left side, next input $4={100}_2$ is given from right side of $5^{th}$ row and the final sum $12+6+2+4=24={0011000}_2$ is computed in $6^{th}$ row. Last row is added to stop the computation.  Note that the example uses $5$ left frame tiles, $4$ corner tiles, $5$ right frame tiles, $7$ top frame tiles, $21$ input tiles and $14$ computational tiles (overall) matching the data given in Table \ref{nsum8tiles}.
%\end{proof}
%Because $\tau = 2,$ only a tile with two neighbors may attach at any time, and therefore, no tile may attach until its right neighbor has.
%The logic of the system is identical to a series of one-bit full adders. Each solution tile takes in a bit from each of the inputs on the south side and a carry bit from the previous solution tile on the east side, and outputs the next carry bit on the west side and the sum on the north side . Because $\tau = 2,$ only a tile with two neighbors may attach at any time, and therefore, no tile may attach until its right neighbor has. After computing the sum of two numbers, we need to combine this sum with next input bitwise. For example if sum is $10010$ and next input is $00010$ the assembly forms $10, 00, 00, 11, 00$ framework. Now it again calculates the sum as explained earlier. This process continues till all the integer inputs are added.

\begin{remark}
We have used $8$ computational tiles in the algorithms, however we can optimize it a bit since the lower portion $( 1 0\;\mbox{or}\;  0 1)$ of computational tiles  gives same effect in computing the sum. Hence while implementing it in "xtilemod" (see Section \ref{xtilemod}) we have used $6$ computational tiles. 
\end{remark}
\begin{table}[htdp]
\caption{Total Tiles used for Computing $\sum_{i=1}^{n} a_i$ using $8$ Computational Tiles}
\begin{center}
\begin{tabular}{|c|c|c|}
\hline
Basic Tiles & Tiles Types & Overall Tiles \\
\hline
Left-Frame & n-1 & 2n-3 \\
\hline
Corner & 4 & 4 \\
\hline
Right-Frame & 2n-3 & 2n-3 \\
\hline
Top Frame & 2 & n+m-1 \\
\hline
Input & (2n-3)(n+m-1) & (n+m-1)(n-1) \\
\hline
Computational & 8 & (n+m-1)(n-2)\\
\hline
\end{tabular}
\end{center}
\label{nsum8tiles}
\end{table}%
\subsection{Addition of $n$ Numbers with $L$ type Tiles}
\begin{theorem}
Let  $\Sigma_2 = \{0, 1, s, ss \} \cup \{\$i | 1 \leq i \leq n+m-1 \} \cup \{\#c | 0 \leq c \leq n \}$ and $T_2$ be a set of tiles (different tiles types of $T_2$ are given in Table \ref{naddLtiles} ) over $\Sigma_2$  as described in Figure \ref{fig:nsumLtiles}.  Then $T_2$ computes the function $\sum_{i=1}^{n} a_i$.
%Let  $S$ be the seed configuration as described in Figure \ref{}
\end{theorem}
The logic of this system is similar to Theorem $2.2$ of \cite{Brun07arith}. The tile set of addition of $n$ numbers using $L$ type tiles of \cite{Brun07arith} have been implemented in xtilemod (see Section \ref{xtilemod}). 
%\begin{proof}
%\end{proof}
\begin{table}[htdp]
\caption{Total Tiles used for Computing $\sum_{i=1}^{n} a_i$ using $L$-type Computational Tiles}
\begin{center}
\begin{tabular}{|c|c|c|}
\hline
Basic Tiles & Tiles Types & Remark \\
\hline
Left-Frame & $1+ \sum_{k=2}^{n} m_k$ &  \\
\hline
Right-Frame & $\sum_{k=2}^{n} m_k$ &  \\
\hline
Top Frame & $2+(n+m-1)$ &  \\
\hline
Computational &  & Bruns' L-Type Tiles\\
\hline
\end{tabular}
\end{center}
\label{naddLtiles}
\end{table}%
\begin{figure}[htdp]
    \label{fig:nsumLtiles}
    \begin{center}

        \subfigure[Typical Computational Tile]{%
            \label{fig:naddLtiles1}
            \includegraphics[width=0.6\textwidth]{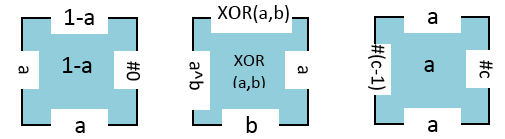}
        }%
       \subfigure[All 8 Computational Tiles]{%
           \label{fig:nsum8tilesb}
           \includegraphics[width=0.50\textwidth]{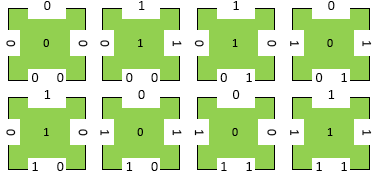}
        }\\ %  ------- End of the first row ----------------------%
        \subfigure[Top Frame Tiles]{%
            \label{fig:naddLtopframetiles}
            \includegraphics[width=0.16\textwidth]{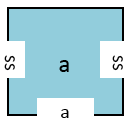}
        }%
          \hspace{.25in}
        \subfigure[Left Frame Tiles \newline ($1 \leq i \leq n-2$)]{%
            \label{fig:naddLlframe}
            \includegraphics[width=0.18\textwidth]{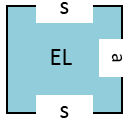}
        }%
        \hspace{.25in}
        \subfigure[Input Frame Tiles \newline $1 \leq i \leq n+m-1$]{%
           \label{fig:naddLinputframe}
           \includegraphics[width=0.2\textwidth]{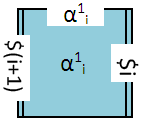}
        }\\ %  ------- End of the first row ----------------------%
        \subfigure[Corner Tiles  ]{%
            \label{fig:addLcorner}
            \includegraphics[width=0.6\textwidth]{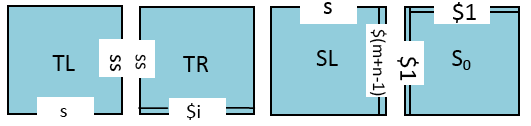}
        }%      
         \subfigure[Example showing addition of $6+4+3+5=18$]{%
            \label{fig:nsum8tilesd}
            \includegraphics[width=0.50\textwidth]{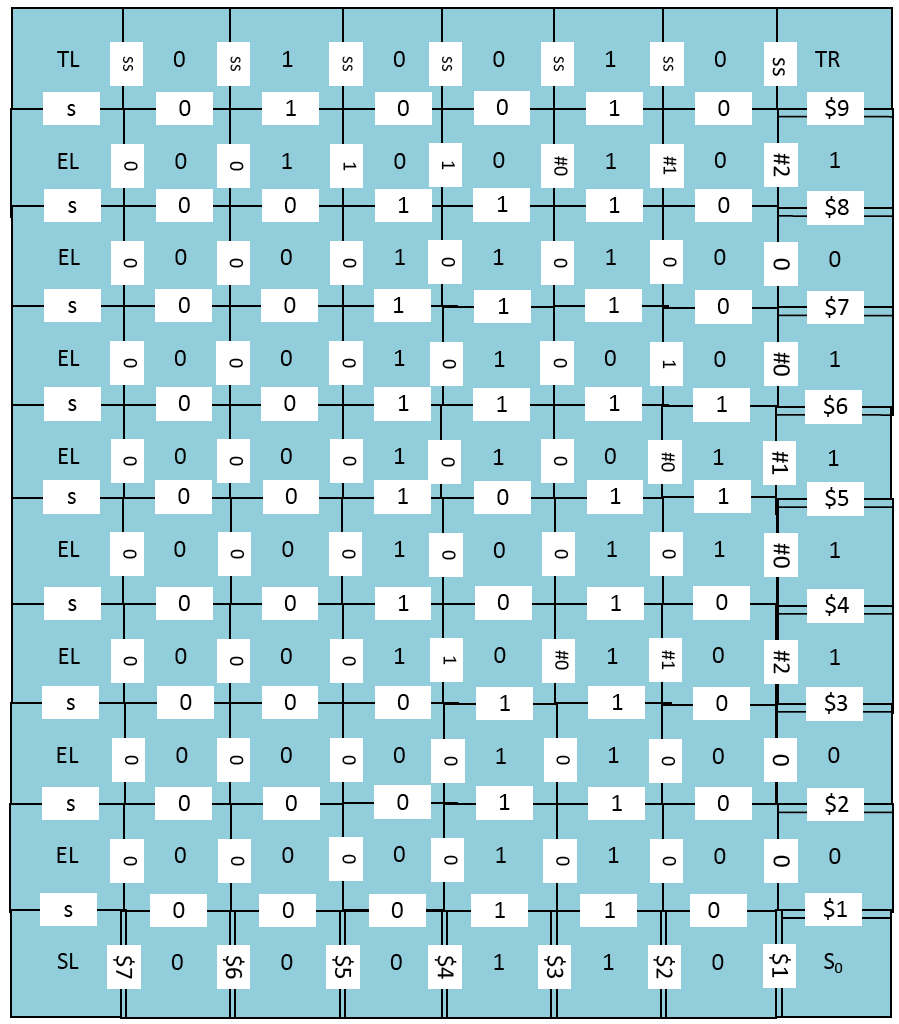}
        }%

    \end{center}
    \caption{%
 Tile Set for Computing  $\sum_{i=1}^{n} a_i$ using $L$ Type Computational Tiles. Note $a,b,c \in \Z_2$.
    }\label{fig:nsumLtiles} %
\end{figure}
\subsection{Multiplication of $n$ Numbers with Tiles}
\begin{figure}[htdp]
   % \label{fig:nmultiply}
    \begin{center}
        \subfigure[Typical Computational Tiles ]{%
            \label{fig:first}
            \includegraphics[width=0.3\textwidth]{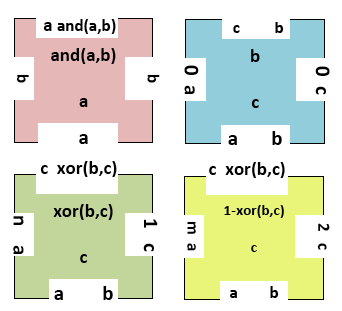}
        }%
        \subfigure[Horizontal Frame Tiles]{%
           \label{fig:second}
           \includegraphics[width=0.3\textwidth]{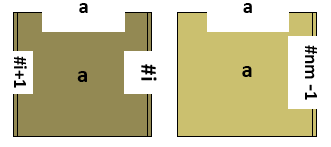}
        }\\ %  ------- End of the first row ----------------------%
        \subfigure[Vertical Frame Tiles]{%
            \label{fig:third}
            \includegraphics[width=0.3\textwidth]{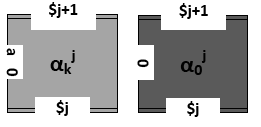}
        }%
        \subfigure[Seed and Result Tiles]{%
            \label{fig:fourth}
            \includegraphics[width=0.5\textwidth]{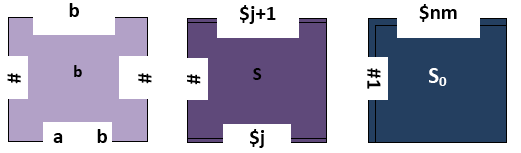}
        }%
    \end{center}
    \caption{%
        Tiles set for Computing $\prod_{i=1}^{n} a_i$. Note $a,b,c \in \Z_2$.
     } \label{fig:nmultiply} %
\end{figure}

\begin{figure}
\centering
\includegraphics[scale=.55]{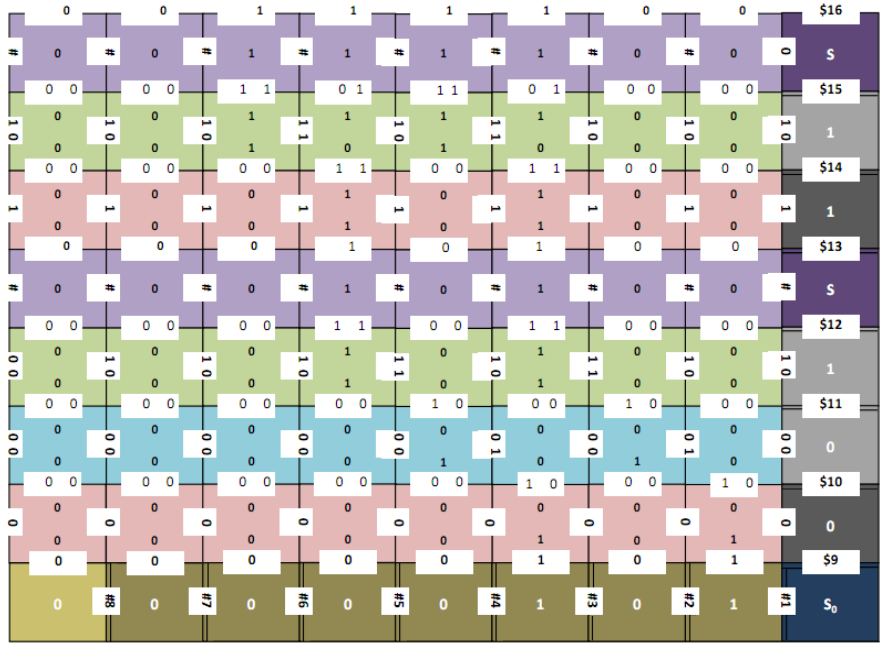}\\
%\centerline{\epsfxsize\hsize\epsfbox{RNA.eps}}
%\includegraphics[scale=.3]{nadderseedLconfig.eps}\\
%\centerline{\epsfxsize\hsize\epsfbox{RNA.eps}}
%\includegraphics[scale=.3]{subtractm.eps}\\
%\centerline{\epsfxsize\hsize\epsfbox{RNA.eps}}
  \caption{Growth of Tiles for Computing the Product of $5 \times 4 \times 3 = 60$}\label{nmultexample}
\end{figure}

\begin{theorem}
Let $\Sigma_3  = \{0, 1, 00, 01, 10, 11, 20, 21, \# \} \cup \{ \# i 1 \leq i \leq nm-2 \} \cup \{\$j: nm \leq j \leq nm +(n-1)+\sum_{k=2}^{n} m_k \}$ and $T_3$ be a set of tiles  (different tiles types of $T_3$ are given in Table \ref{nprod} )  over $\Sigma_3$ as described in Figure \ref{fig:nmultiply}. Then $T_3$ computes the function $\prod_{i=1}^{n} a_i$.
\end{theorem}
%\begin{proof}
Note that the size of $\prod_{i=1}^{n} a_i$ is $n+m$ bits, where $m = \max \{m_j | 1 \leq j \leq n \}$. WLOG, Suppose $a_1$ is of max size  $m$. We put $a_1$ as an input at the bottommost horizontal base tiles 
and all other inputs are given in the vertical frame (right column) followed by a separator tile.  The idea of multiplication of $n$ input integers is a simple extension of the Theorem $2.4$ of  ~\cite{Brun07arith}. The computational tiles also remain the same. The only difference is that $n-2$ inputs are given in the vertical frame separated by one separator tile and subsequently one row. Figure \ref{nmultexample} shows the product of $5 \times 4 \times 3 = 60$. Note that the input $5={101}_2$ is given at the bottom row, input $4={100}_2$ is given as the first vertical input and $3={11}_2$ is given 
as second vertical input. The output is given at the top row as $60={111100}_2$. 
%\end{proof}
\begin{table}[htdp]
\caption{Total Tiles used for Computing $\prod_{i=1}^{n} a_i$}
\begin{center}
\begin{tabular}{|c|c|c|}
\hline
Basic Tiles & Tiles Types & Remark \\
\hline
Horizontal-Frame & $2(nm-1)$ &  \\
\hline
Seed & $1$ &  \\
\hline
Vertical-Frame & $2n+m_j$ &  \\
\hline
Top Frame & $6$ &  \\
\hline
Computational (including a copy tile) & $9$ & \\
\hline
\end{tabular}
\end{center}
\label{nprod}
\end{table}%
\section{Arithmetic Expression of $n$ Numbers with Addition and Subtraction: $\sum_{i=1}^{n} \beta_i a_i,\; \beta_i \in \{+1,-1 \} $}
\begin{figure}[htdp]
  %  \label{fig:nadd_pm}
    \begin{center}
        \subfigure[Typical Computational Tiles for Addition, where \newline $ cout= (a \wedge b) \vee (b \wedge c) \vee (c \wedge a) $ ]{%
            \label{fig:pm1}
            \includegraphics[width=0.3\textwidth]{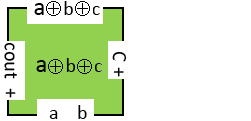}
        }%
            \hspace{.23cm}
        \subfigure[All 8 Computational Tiles for Addition]{%
          \label{fig:pm2}
           \includegraphics[width=0.4\textwidth]{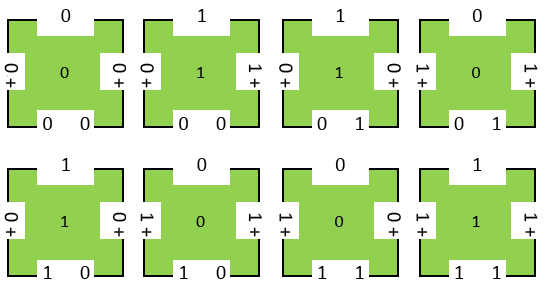}
        }\\ %  ------- End of the first row ----------------------%
        \subfigure[Typical Computational Tiles for Subtraction (Refer Table \ref{bitable})]{%
            \label{fig:pm3}
            \includegraphics[width=0.4\textwidth]{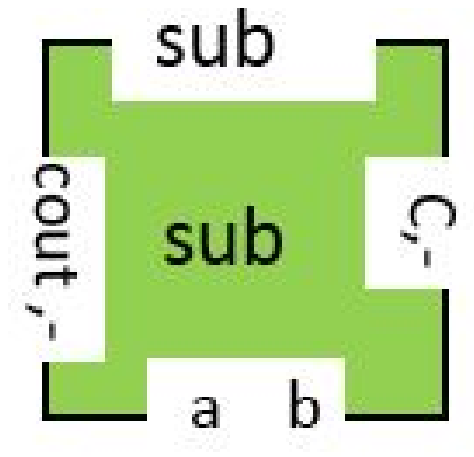}
        }%
          \hspace{.25cm}
        \subfigure[All 8 Computational Tiles for Subtraction]{%
            \label{fig:pm4}
            \includegraphics[width=0.4\textwidth]{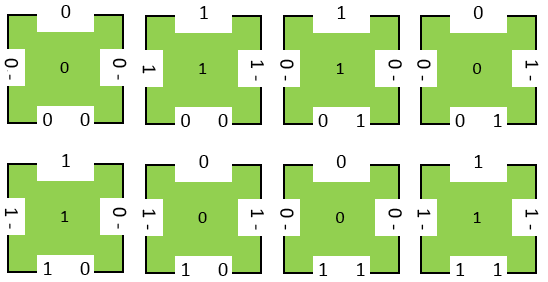}
        }\\%
 \subfigure[Input Tiles for Bottom Row for Two Inputs $1 \leq i \leq m+n-1$]{%
            \label{fig:pm5}
            \includegraphics[width=0.4\textwidth]{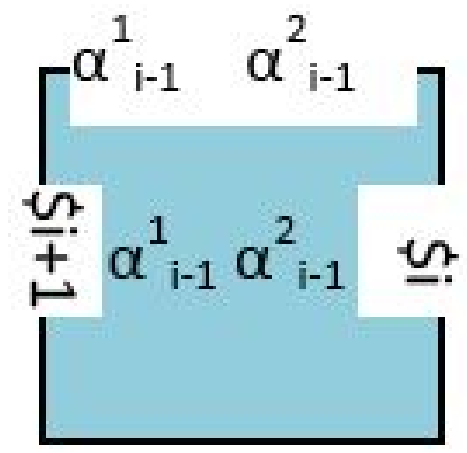}
        }%
        \hspace{.5cm}
 \subfigure[Intermediate Input Tiles ($1 \leq i \leq n+m-1), 3 \leq j \leq n$ $\;\mbox{and}\; \lambda=i+(j-3)(n+m)$  ]{%
            \label{fig:pm6}
            \includegraphics[width=0.4\textwidth]{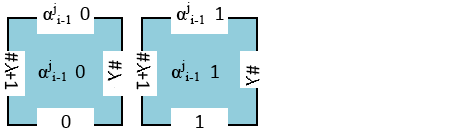}
        }\\%
 \subfigure[Right Frame Tiles 
     ($ER: 1 \leq i \leq n-1)$ \mbox{and} \newline ($SR: 1 \leq i \leq n-2)$ $\;\mbox{and}\; \lambda=1+(i-1)(n+m)$,  $\beta^{i} \in \{+,-\}$ depending on position $i$  ]{%
            \label{fig:pm7}
            \includegraphics[width=0.6\textwidth]{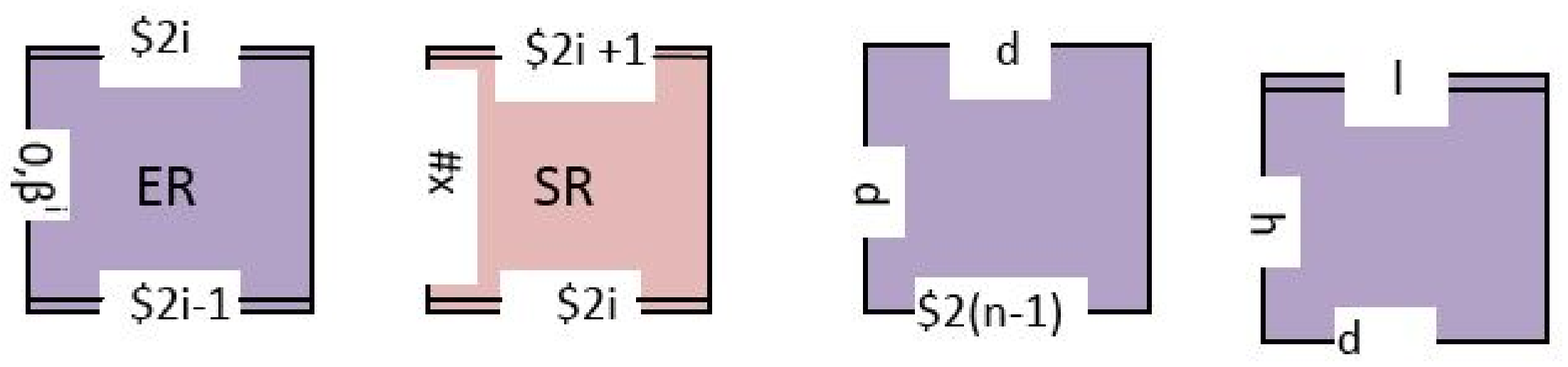}
        }%
 \subfigure[Left Frame Tiles:  $1 < i  < n-2)$, $s \in \{C0,C\_1 \}$ ]{%
            \label{fig:pm8}
            \includegraphics[width=0.49\textwidth]{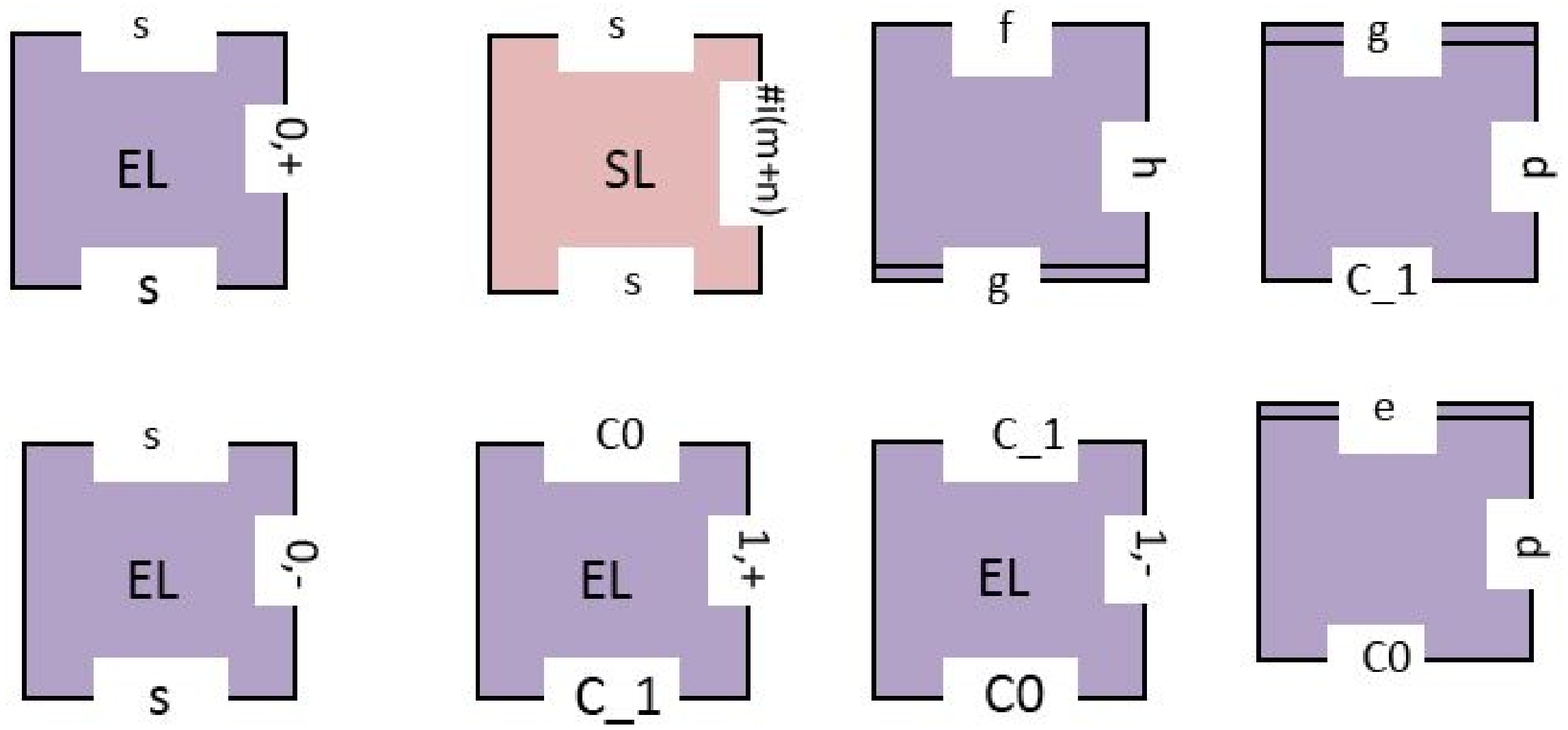}
        }\\% 
 \subfigure[Top Frame Tiles]{%
            \label{fig:pm9}
            \includegraphics[width=0.5\textwidth]{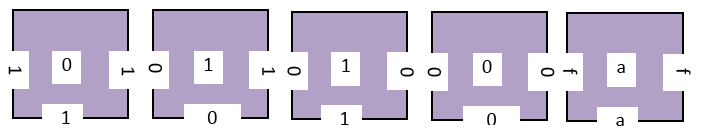}
        }%
\hspace{.25cm}
 \subfigure[Corner Tiles]{%
            \label{fig:pm10}
            \includegraphics[width=0.32\textwidth]{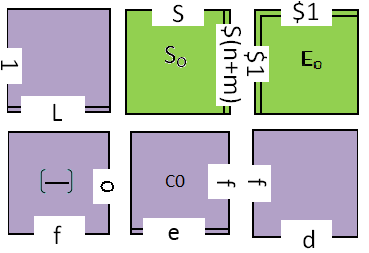}
            }\\%
            \subfigure[Other Computational Tiles]{%
            \label{fig:pm11}     
             \includegraphics[width=0.4\textwidth]{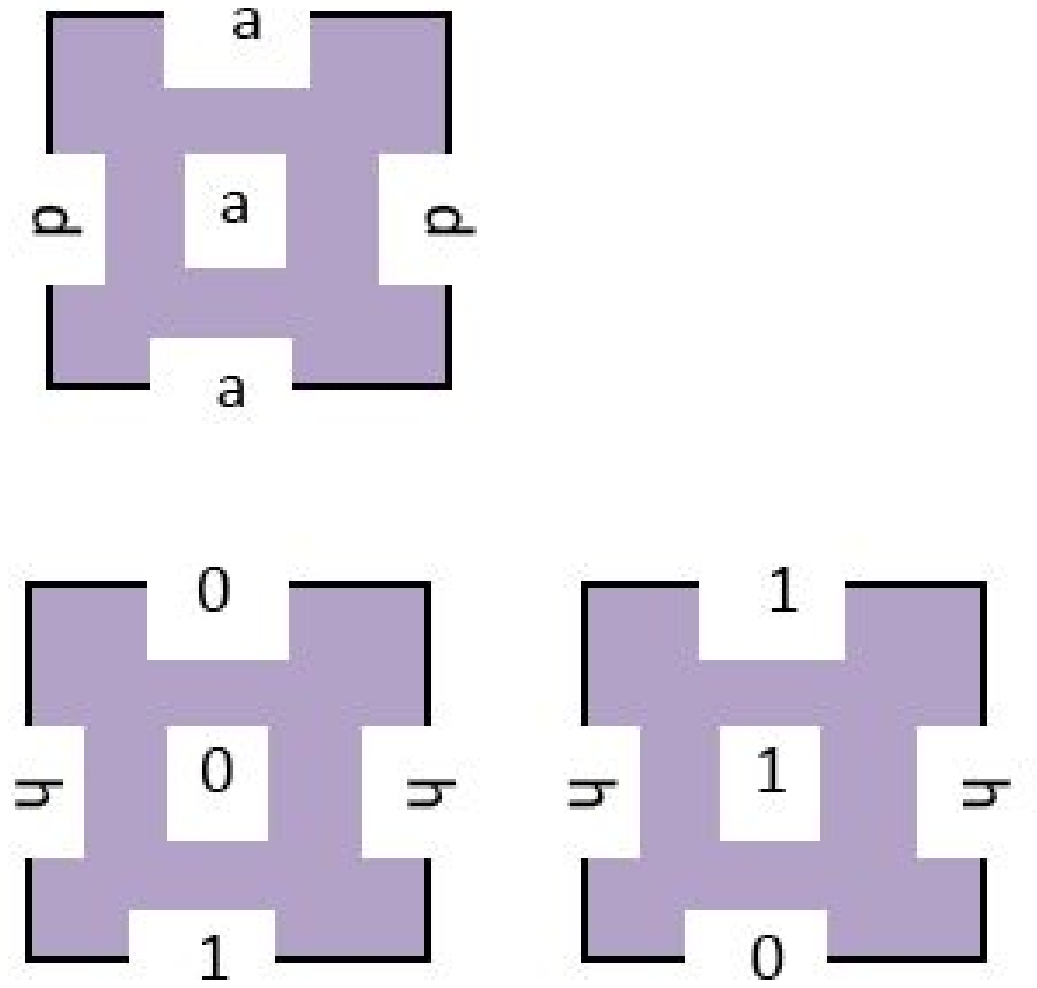}
        }%           
    \end{center}
    \caption{%
    Tile Set for Computing $\sum_{i=1}^{n} \beta_i a_i,\; \beta_i \in \{+1,-1 \},$ Note $a,b,c \in \Z_2$. \label{fig:nadd_pm}
     }%
\end{figure}

\begin{figure}
\centering
\includegraphics[scale=.55]{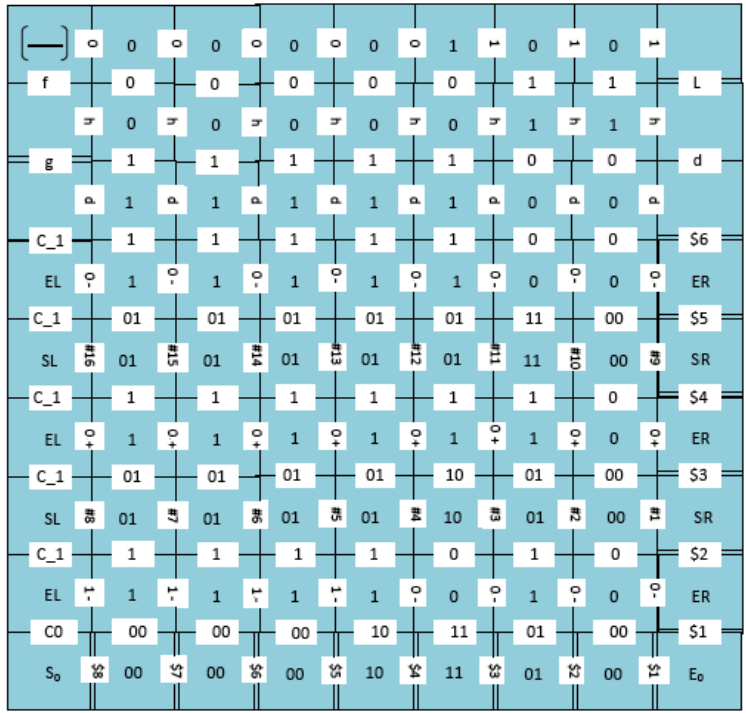}\\
%\centerline{\epsfxsize\hsize\epsfbox{RNA.eps}}
%\includegraphics[scale=.3]{nadderseedLconfig.eps}\\
%\centerline{\epsfxsize\hsize\epsfbox{RNA.eps}}
%\includegraphics[scale=.3]{subtractm.eps}\\
%\centerline{\epsfxsize\hsize\epsfbox{RNA.eps}}
  \caption{Example showing $6-12+4-2=-4$ }\label{nadd_pmexample}
\end{figure}

\begin{theorem}\label{addpmthm}
Let $\Sigma_4  = \{0, 1, 00, 01, 10, 11,  c_{0}, c_{-1} \} \cup \{ d,e,f,g,h,l \}$ be the set of glues and let  $T_4$ be a set of tiles (different tiles types of $T_4$ are given in Table \ref{nplusminus} ) over $\Sigma_4$ as described in Figure \ref{fig:nadd_pm}. Then $T_4$ computes the function $\sum_{i=1}^{n} \beta_i a_i,\; \beta_i \in \{+1,-1 \}$.
\end{theorem}
The logic of the system is identical to a series of one-bit full adders or one bit subtractor depending on the operation which is to be performed. Each solution tile takes in a bit from each of the inputs on the south side and a carry bit from the previous solution tile on the east side, and outputs the next carry bit on the west side and the sum/subtracted value on the north side. In the subtraction the carry bit which is generated on the west indicates the requirement of 1 from the next left tile attached to it. For separating the operation of addition and subtraction + or - sign is attached in all the computational tiles on the east and the west side along with the computational bit. Due to this addition and subtraction can be done by just supplying the sign to computation from the right frame tile. We need to combine this result with next input bitwise. This process continues till all the integer inputs are operated. After operating it if output is positive then the same is displayed on the top row(which is the final output) but if output is negative then the output is converted into unsigned binary bit and this will be displayed on the top row(which is the final output) from the 2Õs complement. This is done because the normal output which was generated for negative value is in the 2Õs complement form. Also a negative sign is displayed with the negative result to tell the user that final output is negative. We will illustrate it by an example. Figure \ref{nadd_pmexample} shows a sample addition of $6-12+4-2= -4.$ Input $6 = {110}_2$  and $12 = {1100}_2$  are given at the bottommost (first) row by adding extra zeros and reading $6$ from right side and $12$ from left side. The subtraction is computed $(6-12 = -6 = {11010}_2)$  in second row and then passed to the $3^{rd}$ row from left side, next input $4 ={100}_2$ is given from right side of $3^{rd}$ row and the sum is computed in $4^{th}$ row  $(6-12+4= -2 = {11110}_2)$ which is then passed to the $5^{th}$ row from left side, next input $2 = {10}_2$ is given from right side of $5^{th}$ row and the final result  $6-12+4-2 = -4 = {1111100}_2$ is computed in $6^{th}$ row. Now as output is negative we have to convert it back into normal unsigned binary expression from $2Õs$ complement with a negative sign in front of it. To this the row $7^{th}, 8^{th}, 9^{th}$ are used. Also the final result ${- 0000100}_2$ is printed on the $9^{th}$ row.
%\begin{proof}
%\end{proof}
%\begin{figure}
%\centering
%\includegraphics[scale=.5]{seedm.eps}
 %\includegraphics[width=0.5\textwidth]{theorem4/naddcorner_pm}
 %\hspace{5cm}
 %\includegraphics[width=0.5\textwidth]{theorem4/naddothercompu_pm}
%\includegraphics[scale=.5]{layeredseedm.eps}\\
%\centerline{\epsfxsize\hsize\epsfbox{RNA.eps}}
  %\caption{$\sum_{i=1}^{n} (-1)^i a_i$  }\label{}
%\end{figure}
%\begin{figure}
%\centering
%\includegraphics[scale=.3]{ctilesadd-m-numbers.eps}\\
%\centerline{\epsfxsize\hsize\epsfbox{RNA.eps}}
%\includegraphics[scale=.3]{addm.eps}\\
%\centerline{\epsfxsize\hsize\epsfbox{RNA.eps}}
%\includegraphics[scale=.3]{subtractm.eps}\\
%\centerline{\epsfxsize\hsize\epsfbox{RNA.eps}}
  %\caption{Computational Tiles for Expression  $\sum_{i=1}^{n} (-1)^i a_i$ }\label{}
%\end{figure}
%\begin{figure}
%\centering
%\includegraphics[scale=.5]{seedm.eps}
%\includegraphics[scale=.5]{layeredseedm.eps}\\
%\centerline{\epsfxsize\hsize\epsfbox{RNA.eps}}
  %\caption{Seed Tiles and Tiles Growth for Expression  $\sum_{i=1}^{n} (-1)^i a_i$  }\label{}
%\end{figure}
\begin{table}[htdp]
\caption{Total Tiles used for Computing $\sum_{i=1}^{n} \beta_i a_i,\; \beta_i \in \{+1,-1 \} $}
\begin{center}
\begin{tabular}{|c|c|c|}
\hline
Basic Tiles & Tiles Types & Remark \\
\hline
Left-Frame & 2n+5 &  \\ %2n+2
\hline
Corner & 6 &  \\
\hline
Right-Frame & 2n-1 &  \\ %2n-3
\hline
Top Frame & 6 &  \\
\hline
Input & (2n-3)(n+m-1) &  \\
\hline
%Computational $\bar{a}+1$& 9 & \\ %9
%\hline
Computational $(+)$ & 8 & \\ %16
\hline
Computational $(-)$ & 8 & \\ %16
\hline
Other Computational & 4 & \\ %16
\hline
\hline
\end{tabular}
\end{center}
\label{nplusminus}
\end{table}%
\section{Modular Arithmetic Expression of $n$ Numbers with Addition and Subtraction: $\sum_{i=1}^{n} \beta_i a_i,\;  \pmod t \;\beta_i \in \{+1,-1 \} $}
\begin{figure}
\centering
\includegraphics[scale=.5]{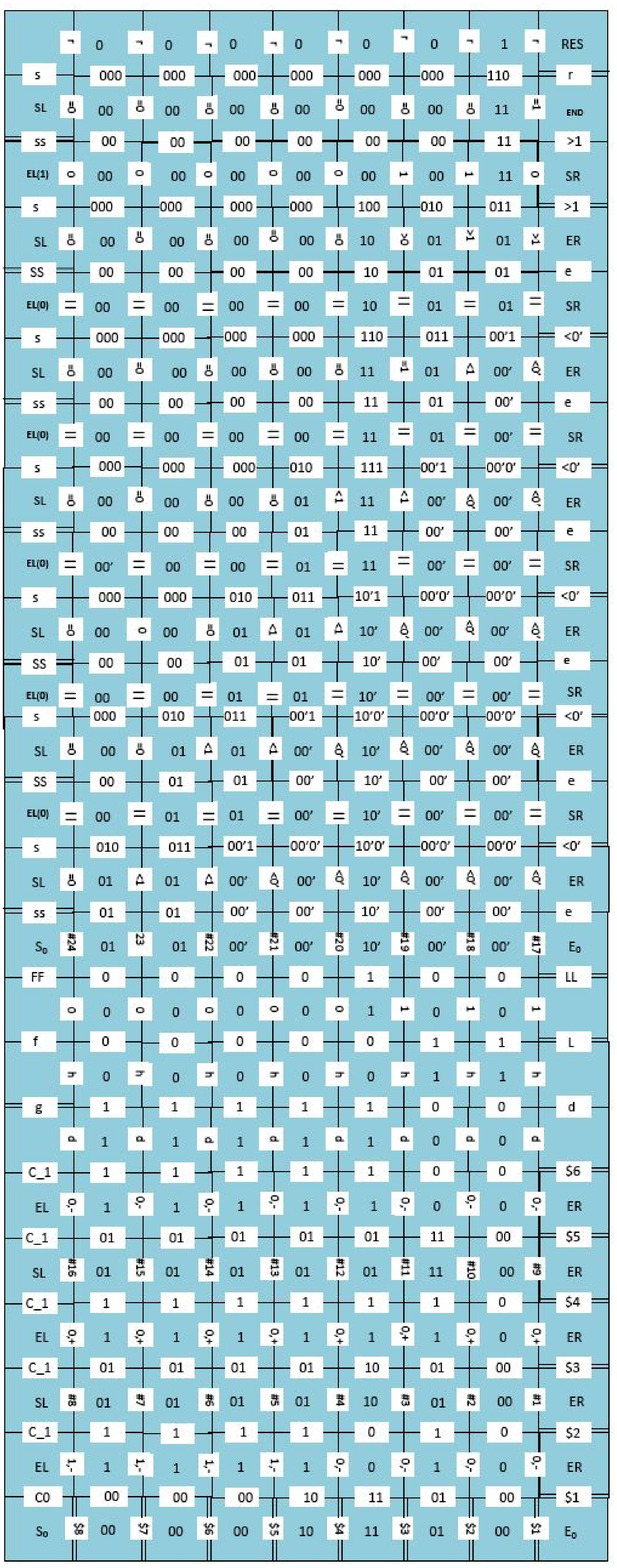}\\
%\centerline{\epsfxsize\hsize\epsfbox{RNA.eps}}
%\includegraphics[scale=.3]{nadderseedLconfig.eps}\\
%\centerline{\epsfxsize\hsize\epsfbox{RNA.eps}}
%\includegraphics[scale=.3]{subtractm.eps}\\
%\centerline{\epsfxsize\hsize\epsfbox{RNA.eps}}
  \caption{Example showing $6-12+4-2=-4$ modulo $3$}\label{naddexample_pmmod}
\end{figure}

%\begin{theorem}
%Let $\Sigma_5  = \{0, 1, 00, 01, 10, 11,  c_{0}, c_{-1} \} \cup \{ d,e,f,g,h,l \}$ be the set of glues and let  $T_5$ be a set of tiles over $\Sigma_5$ as described in Figure \ref{??}. Then $T_5$ computes the function 
%$\sum_{i=1}^{n} \beta_i a_i,\;  \pmod t \;\beta_i \in \{+1,-1 \} $
%\end{theorem}
The tile set for  computing $\sum_{i=1}^{n} \beta_i a_i, \pmod t \;\beta_i \in \{+1,-1 \} $ is just an extended version of the $\sum_{i=1}^{n} \beta_i a_i,\; \beta_i \in \{+1,-1 \} $  tile set (See Theorem \ref{addpmthm}). We add tile set for division (from Lemma $2$ of \cite{Zhang2009377})  to the output of the $\sum_{i=1}^{n} \beta_i a_i,\; \beta_i \in \{+1,-1 \} $ tile set after ignoring sign bit. The final remainder of the output comes on the top most row. An example for $6-12+4-2 \pmod 3=1$ is given in Figure \ref{naddexample_pmmod}. An implementation of the tile set is given in "xtilemod".
%\begin{proof}
%\end{proof}
%\begin{figure}
%\centering
%\includegraphics[scale=.3]{basiccomputationaltilesmodular.eps}\\
%\centerline{\epsfxsize\hsize\epsfbox{RNA.eps}}
%\includegraphics[scale=.3]{addmod.eps}\\
%\centerline{\epsfxsize\hsize\epsfbox{RNA.eps}}
%\includegraphics[scale=.3]{subtractmod.eps}\\
%\centerline{\epsfxsize\hsize\epsfbox{RNA.eps}}
%\includegraphics[scale=.3]{modaddsub_div.eps}\\
  %\caption{Computational Tiles for Expression  $\sum_{i=1}^{n} (-1)^i a_i \pmod n$ }\label{}
%\end{figure}
%\begin{figure}
%\centering
%\includegraphics[scale=.5]{seedmod.eps}
%\includegraphics[scale=.5]{growthseedmod.eps}\\
%\centerline{\epsfxsize\hsize\epsfbox{RNA.eps}}
  %\caption{Seed Tiles and Tiles Growth for Expression  $\sum_{i=1}^{n} (-1)^i a_i\pmod n $  }\label{}
%\end{figure}
\section{Primality Testing with Tiles}
\begin{figure}[htdp]
 %   \label{fig:primetest}
    \begin{center}
        \subfigure[Corner Tiles]{%
            \label{fig:primetest1}
            \includegraphics[width=0.6\textwidth]{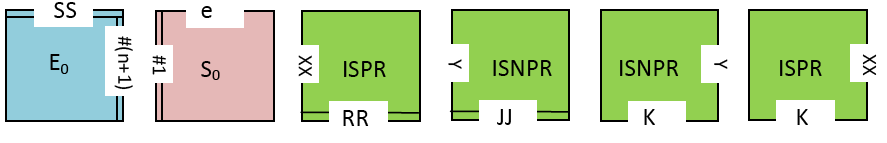}
        }%
        \subfigure[Top Frame Tiles]{%
           \label{fig:primetest2}
           \includegraphics[width=0.40\textwidth]{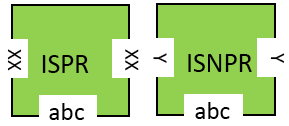}
        }\\ %  ------- End of the first row ----------------------%
        \subfigure[Left Frame Tiles]{%
            \label{fig:primetest3}
            \includegraphics[width=0.5\textwidth]{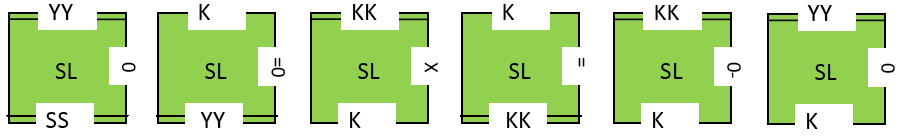}
        }%
          \hspace{.25in}
        \subfigure[Right Frame Tiles]{%
            \label{fig:primetest4}
            \includegraphics[width=0.65\textwidth]{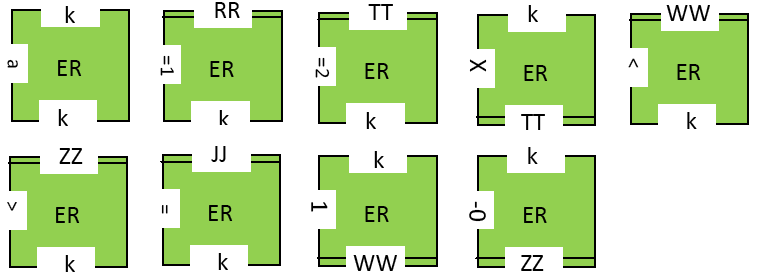}
        }\\%
        \subfigure[Input Tiles \newline 
        ($1 \leq i \leq m$)]{%
            \label{fig:primetest5}
            \includegraphics[width=0.20\textwidth]{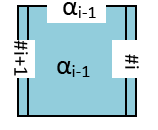}
        }%
        \hspace{.25in}
        \subfigure[Right to Left Move]{%
           \label{fig:primetest6}
           \includegraphics[width=0.20\textwidth]{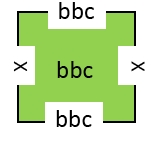}
         }%  
           \subfigure[Subtracting $b-c$ Tiles (Refer Table \ref{bitable})]{%
           \label{fig:primetest7}
           \includegraphics[width=0.20\textwidth]{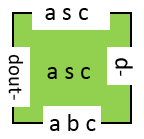}
        }\\ %  ------- End of the first row ----------------------%
      %  \subfigure[Lower Input Tile]{%
          %  \label{fig:primetest7}
        %}%
         \hspace{.3in}
        \subfigure[Subtracting $c-1$ Tiles ]{%
            \label{fig:primetest8}
            \includegraphics[width=0.4\textwidth]{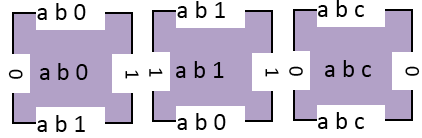}
        }\\% 
         \subfigure[Checking $c > 1$]{%
            \label{fig:primetest9}
            \includegraphics[width=0.5\textwidth]{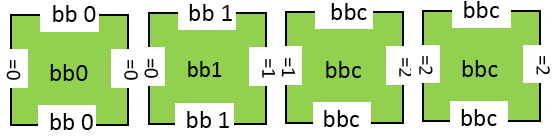}
        }%  
           \subfigure[Checking $c>b$ Out $\in \{=,<, >\} $ depending upon $b=c, b < c, b > c \} $]{%
            \label{fig:primetest10}
            \includegraphics[width=0.5\textwidth]{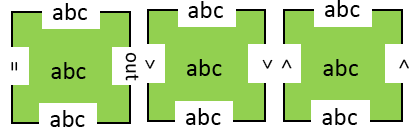}
        }\\%       
         \subfigure[One Bit to Three Bit Converter]{%
            \label{fig:primetest11}
            \includegraphics[width=0.4\textwidth]{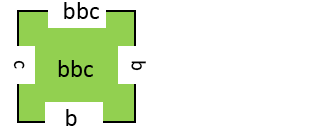}
        }%       
    \end{center}
    \caption{%
        Primality Testing. Note $a,b,c \in \{0,1\}$.
     } \label{fig:primetest} %
\end{figure}

\begin{figure}
\centering
\includegraphics[scale=.5]{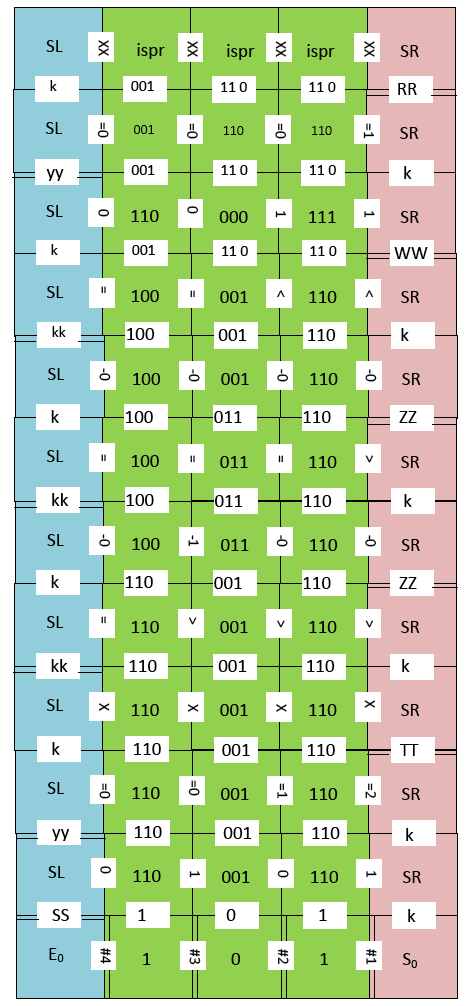}
  \caption{Primality Testing Example for $p=5$ }\label{primetestexample}
\end{figure}

\begin{theorem}
Let $\Sigma_5  = \{0, 1, 000,001,010,011,100,101,110,111 \} \cup \{ =0,=1,=2,0-,1-,=, >, < , SS,RR,JJ,TT,WW,ZZ,KK,YY,XX,Y,K,X \}$ be the set of glues and let  $T_6$ be a set of tiles (different tiles types of $T_5$ are given in Table \ref{primetest} )over $\Sigma_5$ as described in Figure \ref{fig:primetest}. Then $T_5$ determines of primality of any input integer $n$.
\end{theorem}
The logic of primality testing tile set is based on the Algorithm \ref{alg1}.  We will illustrate it by an example. Figure \ref{primetestexample}  shows a sample primality testing for $n=5={101}_2$. This binary input is given at the bottommost (first) row. These bits are copied to left side of the centre of each tile of $3$ bits of second row. Second row also computes $k= \lfloor \frac{n}{2} \rfloor$ and stores it in the right side of the centre of $3$ bits of each tile. The middle bit out of these $3$ bits on each tile centre in the second row is intermediate computation bit $I$ (which is initially the bit of $n$). In fact our input $n$ always remain on left side in each row at the centre of each tile and the value of $k$ remains same till $I-k>0$ is true otherwise it changes to $k-1$ and $I$ becomes $n$. Third row checks if $k >1$, which is true in this case so we copy the content (all $3$ bits) of centre of each tile to $4^{th}$ row (if $k$ is not  $>1$ it quits and displays that it is not a prime). Next row $5^{th}$ checks if the content $I$ corresponding to middle bit is $>k $ (which is true) so $I=I-k$ is done in $6^{th}$ row. Next row further checks if $I-k >0$, which is true so we repeat previous step and $I=I-k$ is done in $8^{th}$ row. Now $9^{th}$ row again checks if  $I-k > 0$ which is not true this time so in $10^{th}$ row we do $k=k-1$ and $I=n$. In $11^{th}$ row again we check if $k >1$ (just like $3^{rd}$ row) and we find that $k=1$ so $12^{th}$ row gives the final output as $5$ is prime. 

\begin{algorithm}                      % enter the algorithm environment
\caption{Primality Testing Algorithm}          % give the algorithm a caption
\label{alg1}                           % and a label for \ref{} commands later in the document
\begin{algorithmic}                    % enter the algorithmic environment
\REQUIRE $n$
\ENSURE $k=\lfloor \frac{n}{2} \rfloor$
\WHILE{$k \neq 1$}
\IF{$n \pmod k  =0$}
\STATE Not Prime
\STATE exit
\ELSE
\STATE $k:=k-1$ 
\ENDIF
\ENDWHILE
\STATE Prime 
\end{algorithmic}
\end{algorithm}

\begin{table}[htdp]
\caption{Total Tiles used for Primality Testing}
\begin{center}
\begin{tabular}{|c|c|c|}
\hline
Basic Tiles & Tiles Types & Remark \\
\hline
Left-Frame & 6 &  \\
\hline
Corner & 6 &  \\
\hline
Right-Frame & 10 &  \\
\hline
Top Frame & 10 &  \\
\hline
Input & n &  \\
\hline
One bit to three bit converter & 4 & \\
\hline
Checking number $>1$ & 12 & \\
\hline
Checking  $c > b$ & 24 & \\
\hline
Right to Left Moving without change & 4 &  \\
\hline
Subtracting  $b-c$ & 16 & \\
\hline
Subtracting  $c-1$ & 16 &  \\
\hline
\end{tabular}
\end{center}
\label{primetest}
\end{table}%

\begin{table}[htdp]
\caption{Subtraction Table Bitwise}
\begin{center}
\begin{tabular}{|c|c|c|c|c|c|}
\hline
a& b & c & sum & cout & Figure \ref{fig:pm3} \\
\hline
b& c & d & s & dout & Figure \ref{fig:primetest7} \\
\hline
 1& 1 &1 &1&1  &\\
\hline
 1& 1 &0 &0&0  &\\
\hline
1& 0 &1 &0&1  &\\
\hline
1& 0 &0 &1&1  &\\
\hline
0& 1 &1 &0&0  &\\
\hline
0& 1 &0 &1&0 &\\
\hline
0& 0 &1 &1&1  &\\
\hline
0& 0 &0 &0& 0& \\
\hline
\end{tabular}
\end{center}
\label{bitable}
\end{table}%
%\begin{proof}
%\end{proof}
\section{Xtilemod: A Software Package for Modular Arithmetic Expressions}\label{xtilemod}
Xtilemod is an modular arithmetic software package implemented in java which generates .tiles file according to options selected at runtime. For example if we have to add two numbers using $8$ tile type addition then we will select the option $1$ which chooses the operation of two numbers and thereafter option $1$ of $1$ which specifies $8$-tile type addition. Last step to get the resultant tile design is to enter the input of integers. The code will display the name of file where the resultant output is loaded. For example if we are adding $6$ and $12$ using $8$ tile type addition the output will be loaded in \begin{verbatim} add_8_tile_12,6.tiles. \end{verbatim} The resultant .tile file can be simulated  using xgrow. The display will be binary. Red color of the resultant tile indicates $1$ and the white indicates $0.$ Resultant tiles are generally at north with a few exception. In case of the division tile set the xgrow compilation displays the quotient at the west and the remainder at the north. In $8$ tile addition of two numbers the output is shown at the middle and the inputs are both in the north and south direction. For more details the reader is referred to the user manual of the software available at http://www.guptalab.org/xtilemod/manual.pdf . The tool 'xtilemod'  is available for download at http://www.guptalab.org/xtilemod. 
%The user manual and online version of the tool is also available on the webpage. 
%%%%%
%\section{\large Software Availability}
%\normalsize The tool 'Xtilemod'  is available for download at http://www.guptalab.org/xtilemod. The user manual and online version of the tool is also available on the webpage. 
%\bigskip
%%%%%
\section{\large Conclusion}
\normalsize In this work we presented tile set for modular arithmetic expressions involving addition, subtraction, multiplication of $n$ integers followed by a modulo operation and a software "xtilemod" for generating the corresponding .tiles files.  A tile set for primality testing is also presented together with a java program "prime.java" for generating .tiles file. Due to space constraints we have omitted the proofs but the key ideas of each result are illustrated  by examples, the detailed proof of the results will be reported in an extended version of the paper elsewhere. \\
{\bf Acknowledgements}\\
\normalsize The authors would like to thank  K. Avinash and B. Bhooma Reddy who have written programs for generating .tiles for addition and subtraction of two input integers.
%%%%%%%%%%%%%%%%%%%%  references  %%%%%%%%%%%%%%%%%%%%%%
\noindent
%%%%%%%%%%%%%%
\bibliographystyle{plain}      % mathematics and physical sciences
\bibliography{dna_xtilemod}   % name your BibTeX data base

\begin{thebibliography}{10}

\bibitem{Brun07fnano}
Yuriy Brun.
\newblock Adding and multiplying in the tile assembly model.
\newblock In {\em Proceedings of the 4th Foundations of Nanoscience:
  Self-Assembled Architectures and Devices ({FNANO}07)}, 2007.

\bibitem{Brun07arith}
Yuriy Brun.
\newblock Arithmetic computation in the tile assembly model: Addition and
  multiplication.
\newblock {\em Theoretical Computer Science}, 378(1):17--31, 2007.

\bibitem{Brun08DNA-lncs}
Yuriy Brun.
\newblock Constant-size tileset for solving an {NP}-complete problem in
  nondeterministic linear time.
\newblock {\em Lecture Notes on Computer Science}, 4848/2008:26--35, 2008.
\newblock A previous version appeared as ``Asymptotically Optimal Program Size
  Complexity for Solving {NP}-Complete Problems in the Tile Assembly Model'' in
  the Proceedings of the 13th International Meeting on {DNA} Computing
  ({DNA}07), pages 231--240, 2007.

\bibitem{Brun08factor}
Yuriy Brun.
\newblock Nondeterministic polynomial time factoring in the tile assembly
  model.
\newblock {\em Theoretical Computer Science}, 395(1):3--23, 2008.
\newblock A previous version appeared as a Center for Software Engineering,
  University of Southern California technical report USC-CSSE-2007-707.

\bibitem{Brun08PhD}
Yuriy Brun.
\newblock {\em Self-Assembly for Discreet, Fault-Tolerant, and Scalable
  Computation on {I}nternet-Sized Distributed Networks}.
\newblock PhD thesis, University of Southern California, 2008.

\bibitem{Brun08np-c}
Yuriy Brun.
\newblock Solving {NP}-complete problems in the tile assembly model.
\newblock {\em Theoretical Computer Science}, 395(1):31--46, 2008.
\newblock A previous version appeared as a Center for Software Engineering,
  University of Southern California technical report USC-CSSE-2007-703.

\bibitem{4656700}
Zhen Cheng, Yufang Huang, and Jin Xu.
\newblock Algorithm for elliptic curve diffie-hellman key exchange based on dna
  tile self-assembly.
\newblock In {\em Bio-Inspired Computing: Theories and Applications, 2008.
  BICTA 2008. 3rd International Conference on}, pages 31 --36, sept. 2008.

\bibitem{WIN2}
Winfree Eric.
\newblock {\em Algorithmic Self Asssembly of {DNA}}.
\newblock PhD thesis, California Institute of Technology, May 1998.

\bibitem{gs87}
B.~Grunbaum and G.~C. Shephard.
\newblock {\em Tilings and Patterns}.
\newblock W.H.Freeman and Company, New York, 1987.

\bibitem{chgo04}
Ho~lin Chen and A.~Goel.
\newblock Error free self-assembly using error prone tiles.
\newblock {\em Tenth International Meeting on DNA Based Computers (DNA10),
  Milano,Italy,June 7-10,2004. Lecture Notes in Computer Science}, 2004.

\bibitem{rsy04}
J.~H. Reif, S.~Sahu, and P.~Yin.
\newblock Compact error-resilient computational dna tiling assembilies.
\newblock {\em Tenth International Meeting on DNA Based Computers (DNA10),
  Milano,Italy,June 7-10,2004. Lecture Notes in Computer Science}, 2004.

\bibitem{rls01}
J.H. Reif, T.H. LaBean, and N.C. Seeman.
\newblock Programmable assembly at the molecular scale: self-assembly of dna
  lattices.
\newblock In {\em Robotics and Automation, 2001. Proceedings 2001 ICRA. IEEE
  International Conference on}, volume~1, pages 966 -- 971 vol.1, 2001.

\bibitem{wang63}
H.~Wang.
\newblock Dominoes and the aea case of the decision problem.
\newblock In {\em Symposiom on Mathematical Theory of Automata}, pages 23--55,
  1963.

\bibitem{win98}
E.~Winfree.
\newblock {\em Algorithmic Self-Assembly of DNA}.
\newblock PhD thesis, California Institute of Technology, Pasadena, CA, 1998.

\bibitem{4281288}
E.~Winfree.
\newblock Algorithmic self-assembly of dna.
\newblock In {\em Microtechnologies in Medicine and Biology, 2006 International
  Conference on}, pages 4 --4, 9-12 2006.

\bibitem{wibe04}
E.~Winfree and R.~Bekbolatov.
\newblock Proofreading tile sets: Error correction for algorithmic
  self-assembly.
\newblock {\em DNA Computers 9, LNCS}, 2943:126--144, 2004.

\bibitem{WIN3}
Eric Winfree.
\newblock The {DNA} and natural algorithms group: the xgrow simulator.

\bibitem{WIN1}
Eric Winfree.
\newblock {DNA} computing by self assembly.
\newblock In {\em NAE's The Bridge}, 2003.

\bibitem{Zhang2009377}
Xuncai Zhang, Yanfeng Wang, Zhihua Chen, Jin Xu, and Guangzhao Cui.
\newblock Arithmetic computation using self-assembly of dna tiles: subtraction
  and division.
\newblock {\em Progress in Natural Science}, 19(3):377 -- 388, 2009.

\end{thebibliography}
%\bibitem{win98}
%E. ~Winfree,
%\newblock {\em Algorithmic Self-Assembly of DNA},
%\newblock Ph. D.  thesis, California Institute of Technology, Pasadena, CA, 1998. 
%\bibitem{xgrow98}
%E. ~Winfree,
%\newblock {\em Xgrow Simulator},
%\newblock Available at http://www.dna.caltech.edu/Xgrow 
%\bibitem{wibe04}
%E. ~Winfree and R. ~Bekbolatov,
%\newblock ``Proofreading tile sets: Error correction for algorithmic
  %self-assembly,''
%\newblock {\em DNA Computers 9, LNCS}, vol.  2943, pp.  126--144, 2004. 
%\bibitem{rsy04}
%J. ~H.  Reif, S. ~Sahu, and P. ~Yin,
%\newblock ``Compact error-resilient computational DNA tiling assemblies,''
%\newblock {\em DNA 10 Italy, June 7-10, 2004.  Lecture Notes in Computer Science},  pp. 293--307, 2004.
%\bibitem{byklwrs00}
%T.H. ~LaBean, H.~Yan, J.~Kopatsch, F.~Liu, E.~Winfree, J. ~H. ~Reif and N. ~C. Seeman,
%\newblock ``The construction, analysis, ligation and self assembly of {DNA} triple crossover complexes,''
%\newblock {\em J.  Am.  Chem.  Soc. }, vol.  122, pp.  1848--1860, 2000. 
%\bibitem{chgo04}
%Ho~lin Chen and A. ~Goel,
%\newblock ``Error free self-assembly using error prone tiles,''
%\newblock {\em Tenth International Meeting on DNA Based Computers (DNA10),
 % Milano, Italy, June 7-10, 2004.  Lecture Notes in Computer Science}, pp. 62--75,  2004.
%\bibitem{wang63}
%H. ~Wang,
%\newblock ``Dominoes and the AEA case of the decision problem,''
%\newblock {\em Symposiom on Mathematical Theory of Automata},  pp. 23--55, 1963. 
%\end{thebibliography}
\end{document}